\documentclass[aps,pre,twocolumn,nokeys,nopacs,floatfix]{revtex4-1}
\usepackage[T1]{fontenc}
\usepackage[colorlinks=true,hyperfootnotes=true,breaklinks=true]{hyperref}
\usepackage{auto-pst-pdf}
\usepackage{graphicx}
\usepackage{psfrag}

\begin{document}
\title{Mental ability and common sense in an artificial society}
\author{Krzysztof Malarz}
\homepage{http://home.agh.edu.pl/malarz/}
\email{malarz@agh.edu.pl}
\author{Krzysztof Ku{\l}akowski}
\email{kulakowski@fis.agh.edu.pl}
\affiliation{\href{http://www.agh.edu.pl/}{AGH University},
\href{http://www.pacs.agh.edu.pl/}{Faculty of Physics and Applied Computer Science},\\
al. Mickiewicza 30, 30-059 Krakow, Poland.}

\begin{abstract}
Having equally valid premises {\em pro} and {\em contra}, what does a rational human being prefer? 
The answer is: nothing. 
We designed a test of this kind and applied it to an artificial society, characterized by a given level of mental ability. 
A stream of messages from media is supplemented by ongoing interpersonal communication. 
The result is that high ability leads to well-balanced opinions, while low ability produces extreme opinions.  
\end{abstract}

\date{\today}
\maketitle

\section{How we decide what is right or wrong}

We read newspapers and watch TV every day. There are many issues and many controversies. 
Since media are free, we can hear arguments from every possible side. 
How do we decide what is wrong or right? 
The first condition to accept a message is to understand it; messages that are too sophisticated are ignored. 
So it seems reasonable to assume that our understanding depends on our ability and our current knowledge. 
Here we show that the consequences of this statement are surprising and funny.

\section{How do we learn}

To demonstrate this, we propose a computational model with two assumptions ~\cite{1}. 
The first is that messages can be represented as points on a plane of a finite area, say, a square $a\times a$. 
A consequence is that we can measure the distance between messages. 
The second assumption is that we can understand a message if it is not too far from what we already know.

As a direct consequence of these two assumptions, we obtain a simple model of learning.
In this model the mind is represented by an area around the messages understood by the mind's owner. 
Her/his ability is represented by a critical distance $D_c$.
A new message can be grasped if its distance to the closest previously understood message is shorter than $D_c$.
If this distance is longer, the message is ignored.

\begin{figure}[htp]
\centering
a) \includegraphics[width=1\columnwidth]{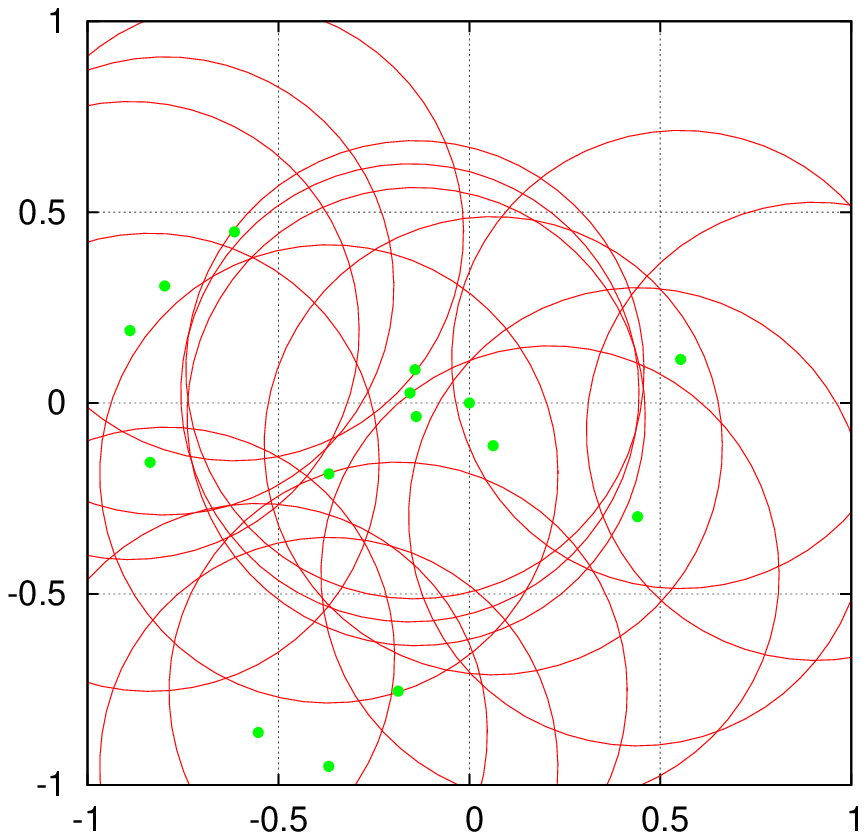}\\
b) \includegraphics[width=1\columnwidth]{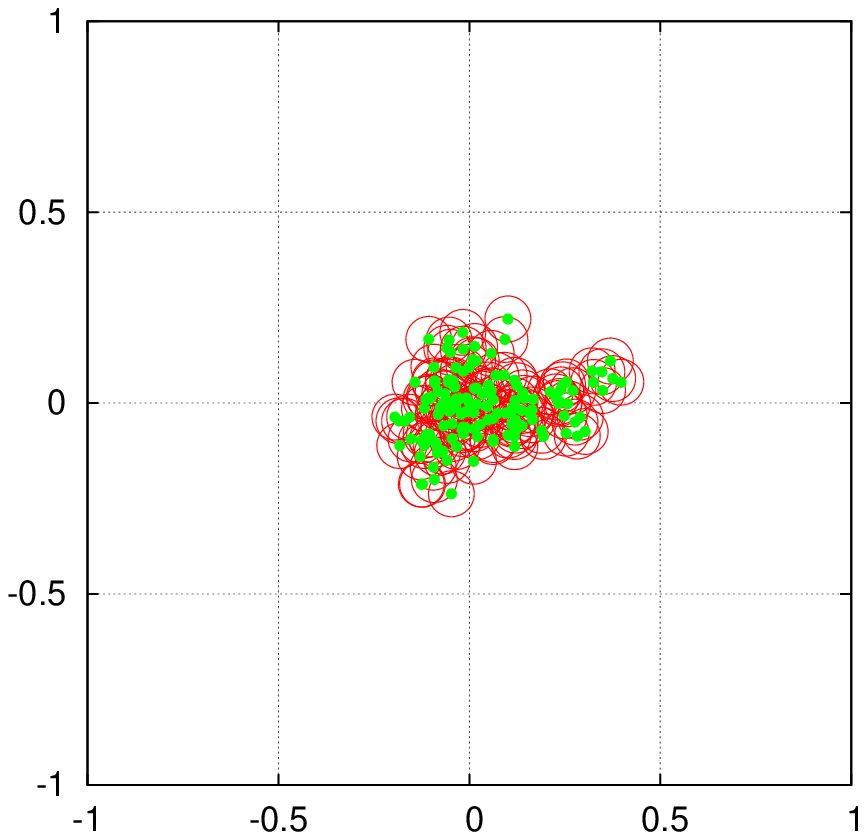}
\caption{\label{fig:1} The `mental history' of a single actor: positions of understood messages for a) $D_c/a = 0.3$ and b) $D_c/a = 0.03$. Each actor starts from one message at the center of the square. In case a) the actor understands almost all messages after a few steps. In case b) the actor remains confined with her/his knowledge, with a bias towards right (the bias direction is random).}
\end{figure}

\begin{figure}[htp]
 \centering
 \psfrag{D_c/a}{$D_c/a=$}
 \psfrag{0.03}{$0.03$}
 \psfrag{0.3}{$0.3$}
 \includegraphics[width=\columnwidth]{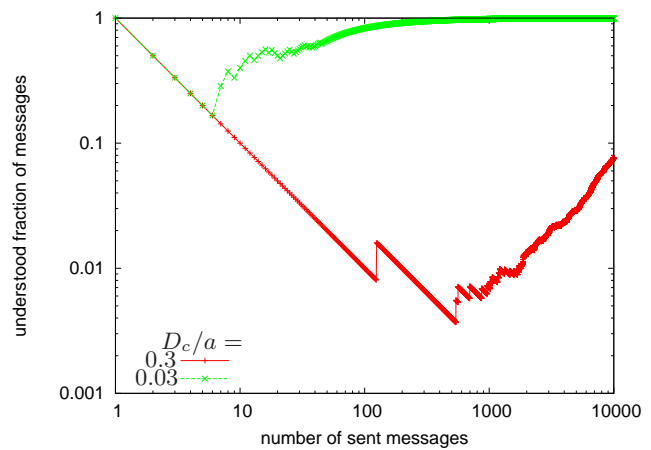}
 \caption{\label{fig:2} The fraction of messages understood as a function of the number of incoming messages, for $Dc/a=0.3$ (upper, green curve) and $Dc/a=0.03$ (lower, red curve). Each actor starts from one message at the center of the square. In this log-log plot, the lower curve shows a vertical step between two subsequently understood messages. Initially, such events are rather rare.}
\end{figure}

Let us consider a new area of experience: differential calculus, traffic regulations, stock market, foreign policy or classic Latin grammar can serve as examples.
Initially we know a small area on a square.
Step by step, we expand our knowledge each time when a new message is found to be comprehensible.
The speed at which the known area expands is determined by the parameter $D_c$.
If it is comparable with size a of the square, the mind understands everything after a few messages.
However, if  $D_c/a$ is small, initially most of the incoming messages are ignored, and the area of understanding expands very slowly.
This is demonstrated in Fig.~\ref{fig:1}, where we see an area equivalent to the gained knowledge for $D_c/a =0.3$ after 10 messages in panel a), and for $D_c/a =0.03$ after 100 messages in panel b).
In Fig.~\ref{fig:2} we show the fraction of messages that are understood as a function of the number of all messages, also for these two values of $D_c/a$.

\section{An example---how we think about politics}

As we are political animals, let us apply the model to our political beliefs. In this field, public discussions are most aggressive and arguments least convincing.
Trying to be objective---as scientists should be---we choose our square to be symmetrically divided between two orientations: Left and Right.
The vertical axis can be interpreted as a measure of the distance between Authoritarian and Libertarian, as in~\cite{2}.

Let us suppose that our model mind is target of a stream of messages, evenly distributed in the square.
Again, if $D_c/a$ is close to 1, the situation is rather trivial; the model mind quickly arrives at full understanding.
However, for small values of the ratio $D_c/a$ the situation is less trivial because bias comes into play. Let us assume at first that our mind is initially unbiased; its owner accepts the first message if it appears within a circle of radius $D_c$ around the square centre.
Yet we can expect that a certain degree of bias will soon develop: it is unlikely that the Left-Right symmetry is preserved for the trajectory of a single mind.
An example of this effect is shown clearly in Fig.~\ref{fig:1}b.

To explain this, we refer to theory of random walk~\cite{3}. Suppose that our model is simplified to one dimension, with steps towards left and right at discrete times with equal probabilities.
Suppose also that our mind made a step in a given direction. We can then ask the question: how long will it take until it returns?
The theory tells us: infinitely long on average.

Here we touch upon an important feature of our model.
It is clear that each mind will reach full understanding after a sufficiently long time.
However, the difference between able (large $D_c$) and less able (small $D_c$) minds manifests itself always within a finite time.
It is just our own lifetime which is finite, and its length provides a time scale for everything we do, including understanding things.
Compared with the 'infinitely long' case from the previous paragraph, this means that, once biased, many of us will never reach objectivity again.

\section{The test}

To generalize our model, let us consider a large number of minds, which are again target of a stream of messages. As we have seen, whether we include their initial bias or not is of secondary importance.
Now we are going to design a test of social common sense in our artificial society.
As the messages are evenly distributed, neither Left nor Right arguments prevail.
Knowing this, we can expect that a reasonable person remains objective. What is the result?

To answer this, let us introduce a probability $p$ that a given mind's owner, when asked about her/his preference, is going to answer `Right'.
Likewise, a probability $q$ is assigned to the answer `Left', with the obvious condition $p+q=1$.
For each mind, the probability $p$ will be calculated as follows.
The number of all messages he or she understood within a given time is $N$.
This set is divided into $N(L)$ and $N(R)$, where $N(L)$ is the number of understood messages placed on the left part of the square, and N(R) for the right part.
Obviously, $N(L)+N(R)=N$. Then, $p=\langle x(R)\rangle/\langle|x|\rangle$, where  
  \begin{equation}
  \langle x(R)\rangle=\sum_i^{N(R)} x(i)
  \end{equation}
is the mean $x$-coordinate of messages on the right half-plane, and, 
  \begin{equation}
   \langle |x|\rangle=\sum_j^N |x(j)|
  \end{equation}
is the mean absolute value of the $x$-coordinate of all messages.
Probability $p$ is now calculated separately for each mind \cite{1}.

\begin{figure}[htp]
 \centering
 \psfrag{H(p)}{$H(p)$}
 \psfrag{p}{$p$}
 \psfrag{D_c/a}{$D_c/a=$}
 \psfrag{0.03}{$0.03$}
 \psfrag{0.3}{$0.3$}
 \includegraphics[width=\columnwidth]{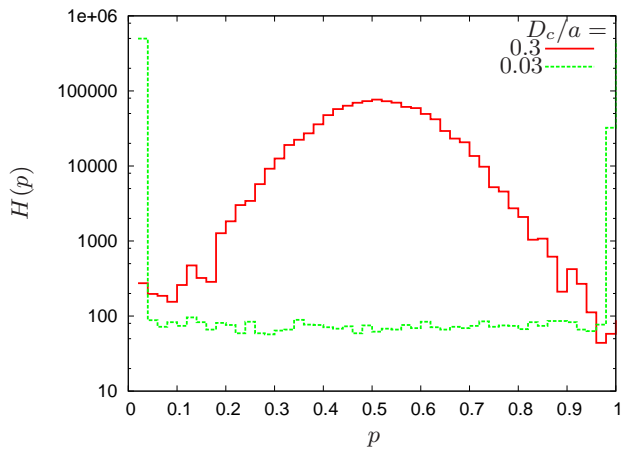}
 \caption{\label{fig:3} Histogram $H(p)$ of the individual probabilities $p$ for $D_c/a = 0.3$ (central peak) and $D_c/a = 0.03$ (binomial curve). 
The small asymmetry of the latter is a statistical fluctuation. 
The results are averaged over $10^3$ actors, 100 messages and $10^3$ runs; one run means one set of messages, the same for all actors. 
Here, initial positions of actors (their first understood messages) are evenly distributed on the square.}
\end{figure}

What is the probability distribution of $p$ itself? The answer is shown in Fig.~\ref{fig:3}, for different values of the ability parameter $D_c$.
As we see, both plots preserve the Left-Right symmetry within the accuracy of statistical errors. For large values of $D_c$ the resulting probability distribution is centred around the value $p=\frac{1}{2}$.
By contrast, for small $D_c$ the distribution consists of two sharp maxima close to $p = 0$ and $p = 1$.
In other words, in the former case of large ability a statistical mind answers `Left' and `Right' with equal probabilities.
This is equivalent to the answer ``I don't know'', the only reasonable answer because the incoming messages do not provide arguments for a more decisive statement.
However---and this is our most important result---for a society characterized with small ability $D_c$ a statistical mind answers either surely `Left', or surely `Right'.
In other words, in the case of small mental ability all opinions are extreme.

\section{Additional remarks}

The model \cite{1} has been further developed to include consequences of interpersonal communication: minds not only hear but also articulate their opinions, which are included to the stream of messages.
To mention two main results, we note that an intensive communication leads to a clustering of opinions, which become more extreme even for the case of moderate ability \cite{4}.
On the other hand, the latter unanimity disappears if messages are addressed to minds which are neighbours in the square of issues.
Then, again, the opinions are less extreme \cite{5}.
These results lead one to be cautious about situations in which unanimity is treated as good and conflict as evil. Alas, in our world unanimity is almost always against somebody else.
In that case the contradistinction is not `unanimity vs. conflict', but rather `diversity vs. extreme'.

Paraphrasing Paul G\'eraldy, one could say that it is the political party who chooses the man who will choose her.
This means that everybody will be chosen by some party.
Yet a simple ``I don't know'' seems a good remedy against an extreme `Yes' or an extreme `No'.
What is funny (at least for us) is that this is the result of a model based on statistical mechanics.

\section*{About the Authors}
K.~K. (born 1952) and K.~M. (born 1972) have been working together at AGH University, Krakow, for some 15 years on networks, sociophysics and other interdisciplinary applications of statistical mechanics. 
K. K. (Ph.D. in Physics, full professor) teaches nonlinear dynamics and game theory, K.~M. (Ph.D. in Physics, assistant professor) teaches cellular automata.
Both are members of the Polish Physical Society, while K.~K. is also member of the Polish Sociological Society.
K.~M. is an individual member of the European Physical Society as well.

\end{document}